\documentclass{emulateapj}
\usepackage{times}
\usepackage{aas_macros}
\usepackage{graphicx}
\usepackage{courier}
\usepackage[T1]{fontenc}
\usepackage{array}
\usepackage{textcomp}
\usepackage{multirow}
\usepackage{footnote}
\usepackage{epstopdf}
\newcommand{\degrees}{\ifmmode^{\circ}\else$^{\circ}$\fi}
\newcommand{\amin}{\ifmmode^{\prime}\else$^{\prime}$\fi}
\newcommand{\asec}{\ifmmode^{\prime\prime}\else$^{\prime\prime}$\fi}

\def\lsim{\mathrel{\rlap{\lower4pt\hbox{\hskip1pt$\sim$}}
    \raise1pt\hbox{$<$}}}                
\def\gsim{\mathrel{\rlap{\lower4pt\hbox{\hskip1pt$\sim$}}
    \raise1pt\hbox{$>$}}}                

\shorttitle{Millisecond Pulsar Companions}
\shortauthors{Schroeder \& HalpernTable}

\begin{document}

\title{Observations and Modeling of the Companions of Short Period Binary Millisecond Pulsars: Evidence for  High-Mass Neutron Stars.}

\author{Joshua Schroeder\altaffilmark{1}\altaffilmark{2}, Jules Halpern\altaffilmark{1}}
\altaffiltext{1}{Department of Astronomy, Columbia University, Mail Code 5246, 550 West 120th Street, New York, NY 10027, USA}
\altaffiltext{2}{Harvard-Smithsonian Center for Astrophysics, 60 Garden Street, Cambridge, MA 02138, USA}

\begin{abstract}

We present observations of fields containing eight recently discovered binary millisecond pulsars using the telescopes at MDM Observatory. Optical counterparts to four of these systems are detected, one of which, PSR J2214+3000, is a novel detection. Additionally, we present the fully phase-resolved B, V, and R light curves of the optical counterparts to two objects, PSR J1810+1744 and PSR J2215+5135 for which we employ model fitting using the ELC model of \citet{Orosz2000} to measure the unknown system parameters. For PSR J1810+1744 we find that the system parameters cannot be fit even assuming that 100\% of the spin-down luminosity of the pulsar is irradiating the secondary, and so radial velocity measurements of this object will be required for the complete solution. However, PSR J2215+5135 exhibits light curves that are extremely well constrained using the ELC model and we find that the mass of the neutron star is constrained by these and the radio observations to be $M_{\rm NS} > 1.75 M_\Sun$ at the $3-\sigma$ level. We also find a discrepancy between the model temperature and the measured colors of this object which we interpret as possible evidence for an additional high-temperature source such as a quiescent disk. Given this and the fact that PSR J2214+5135 contains a relatively high mass companion ($M_{\rm c} > 0.1 M_\Sun$), we propose that similar to the binary pulsar systems PSR J1023+0038 and IGR J18245-2452, the pulsar may transition between accretion- and rotation-powered modes.

\end{abstract}


\section{Introduction}
\label{sec:Introduction}

Millisecond pulsars have been a source of fascination and speculation since their discovery in 1982 by \citet{Backer1982}. Shortly thereafter, \citet{Alpar1982} and \citet{RS1982} offered a theoretical explanation for their existence, namely that Low-Mass X-ray Binaries (LMXBs) and isolated millisecond pulsars were connected by a recycling evolutionary scenario where, during the LMXB phase, the pulsar was spun-up and then the companion was eliminated through an ablation or merger process. Support for this evolutionary scenario came with the identification of the ``Black Widow Pulsar'', PSR B1957+20, which was discovered by \citet{FruchterStinebringTaylor1988} and now serves as the prototype for an entire class of binary millisecond pulsars in tight orbits ($P_{\rm orb} \le 1\ {\rm d}$) with low mass ($M_{\rm c} \le 1\ M_{\Sun}$) companions. Such systems are characterized by radio eclipses around the orbital phase $\phi = 0.25$ (where the ascending node corresponds to $\phi = 0$) consistent with a high plasma-density environment when the line-of-sight to the pulsar is occulted by the environment surrounding the companion. This feature has been cited in the case of the original Black Widow Pulsar as evidence that the companion is being ablated by action from either high-energy radiation from the pulsar or by the pulsar's particle wind \citep{Ruderman1989}. Shortly after the discovery of the radio eclipses, \citet{Fruchter1988} discovered the optical counterpart to B1957+20 whose light curve had orbital modulation consistent with an irradiation heating mechanism and \citet{Kluzniak1988} proposed that the companion was being ablated thus providing an explanation for both the radio and optical data, the necessary high plasma density being due to material evaporated from the companion.

Until the launch of the Fermi Gamma Ray Space Telescope, only three black widow-type pulsars outside of globular clusters were known to exist, B1957+20, J2051-0827 \citep{Stappers1996}, and J0610-2100 \citep{Burgay2006} -- the paucity of such sources attributed to the computational difficulty of discovering millisecond pulsars in blind radio surveys. Millisecond pulsars, however, are strong gamma-ray sources, and so the technique of searching for radio-loud pulsars in the error box of an unassociated gamma-ray source discovered by the Fermi Large Area Telescope (LAT) has been fruitful. As of writing, the Fermi team has identified gamma-ray sources associated with 51 millisecond pulsars of which 39 have been determined to be in binary systems, and all but one of these Fermi-identified pulsars have been associated with radio counterparts \citep{FermiCat2013}. 

Radio observations of binary millisecond pulsars allow for the identification of eclipse features, while ephemeris fits precisely determine certain system parameters including the position on the sky, binary period ($P_{\rm orb}$), epoch of the ascending node ($T_0$), and projected semi-major axis ($x$) \citep{Ray2012}. These values can be related to the physical characteristics of the systems through the binary mass function 
\begin{equation}
f = \frac{4 \pi^2 x^3}{G P_{\rm orb}^2} = \frac{\left(M_{\rm c} \sin i\right)^3}{\left(M_{\rm NS} + M_{\rm. c}\right)^2} \label{eq:mass}
\end{equation}
where $M_{\rm NS}$ is the mass of the neutron star, $M_{\rm c}$ is the mass of the companion, and $i$ is the inclination angle of the orbit. The determination of masses for the objects in the system requires measuring the inclination angle and one of the object masses or the mass ratio ($Q \equiv M_{\rm NS}/M_{\rm c}$) \citep{Lorimer2004}.

Following the discovery of many new short period binary millisecond pulsars, \citet{Roberts2011} proposed that a separate class of radio-eclipsing binaries he called ``redbacks'' exist with binary mass functions in excess of $f \ge 10^{-3} M_{\odot}$, considerably larger than they typical black widows, which, as a class, have mass functions closer to $f \sim 10^{-5} M_{\odot}$. Radio eclipse properties of redback systems are qualitatively different than those of canonical black widow pulsars; they exhibit longer eclipses that are more irregular and variable in their phase of onset and duration (Ransom, personal communication).  Extrapolating the distinction between the subpopulations to plausible masses for the millisecond pulsars ($M_{\rm NS} \sim 1.4 M_{\odot}$) implies that redback companions are of main sequence or subdwarf masses ($M_{\rm c} \gsim 0.2 M_{\odot}$) while the black widow companions are the masses of brown dwarfs or stripped white dwarf cores similar to those seen in ultracompact LMXBs \citep{Rappaport1982, Deloye2003}. Short period millisecond pulsars with redback-sized masses include PSR J1023+0038, a ``missing link" pulsar that switched from a LMXB accreting system to a millisecond pulsar binary within the last decade \citep{Archibald2009}, while another example of a transitioning pulsar, IGR J18245-2452, was recently identified in the globular cluster M28 as switching from being a rotation powered to being an accretion powered pulsar \citep{Papito2013}.

In the last few years, optical observations of binary millisecond pulsar systems have been used to great effect as a means to constrain system parameters including the mass of the neutron stars. In particular, \citet{vanKerkwijk2011} obtained a constraint of $M_{\rm NS} > 1.9\ M_{\odot}$ for B1957+20 while \citet{Romani2012} used photometric and spectroscopic measurements of J1311-3430 to obtain a lower mass limit of the neutron star of $M_{\rm NS} > 2.1\ M_{\odot}$. Recently, \citet{Kaplan2013} used a radial velocity measurements of the optical companion of J1816+4510 to obtain an inclination-angle dependent neutron star mass of $M_{\rm NS} \sin^3 i = 1.84 \pm 0.11\ M_\Sun$. Similarly \citet{Crawford2013} detected the companion to the redback J1723-2837 and used spectroscopic follow-up to obtain an inclination-angle dependent neutron star mass of $M_{\rm NS} \sin^3 i = 0.3 \pm 0.1 M_\Sun$ from which they predict an inclination angle for the system of $i \le 41\degrees$.

\citet{Ray2012} provides the Fermi LAT team's list of radio-identified binary millisecond pulsars for which various radio-measured properties have been obtained including the pulsar spin periods, binary mass functions, orbital periods, and dispersion measures. Of particular interest for optical follow-up campaigns is the positional accuracy to the sub-arcsecond level determined from the radio ephemerides which is a great improvement over the ten arcminute-scale error boxes associated with the LAT confidence regions. After detecting the companion object, it is possible to constrain the inclination angle and the blackbody temperature profile of the secondary through photometry alone. An example of such a study is one done by \citet{Breton2013} who identified the optical counterparts of four Fermi-detected binary millisecond pulsars and measured the magnitudes in various filters at a number of phases over each orbit. It has generally been assumed that meaningful constraints on masses would require radial velocity data, but photometrically stable and relatively high signal-to-noise light curves can give model constraints on the masses of the primary and secondary, especially if ellipsoidal variations are detected. Using photometry alone to obtain mass constraints was first suggested in foundational work by \citet{Avni1975} and a similar technique was used by \citet{Jackson2012} to constrain the mass of exoplanets.

In this work, we report on observations made over the course of seven runs at MDM Observatory of the fields containing eight of the Fermi-detected binary millisecond pulsars with periods less than one day. Of the eight systems, we detect optical counterparts of four, three of which have been reported in the literature and one of which is a novel detection. Two of these four were bright enough to allow us to obtain phase-resolved optical light curves in BVR filters while detection upper limits were made for the others. In Section~\ref{sec:Observations and Data Analysis} we describe the observations and the data reduction procedures including the positive identifications and upper limits of optical counterpart detections, Section~\ref{sec:Modeling and Parameter Estimation} makes use of the ELC code \citep{Orosz2000} modeling to do parameter fitting for two of the best-observed objects, and Section~\ref{sec:Discussion} is a discussion of possible interpretations and implications of our results while Section~\ref{sec:Conclusions} gives the conclusions of the paper.

\section{Observations and Data Analysis}
\label{sec:Observations and Data Analysis}

Over the course of seven observing runs at MDM observatory from May 2010 to August 2011, we observed the positions of various Fermi-detected and radio-confirmed binary millisecond pulsars using the 1.3 meter McGraw-Hill and the 2.4 meter Hiltner telescopes. For the first six runs, we employed one of either two thinned backside illuminated (a $2048 \times 2048$ pixel CCD called "Echelle" and a $1024 \times 1024$ pixel CCD called "Templeton")  or, one thick frontside illuminated imaging CCD (a $2048 \times 2048$ pixel CCD called "Nellie") with Harris B, V, and R filters. The MDM CCD-control software failed for our last two runs in August 2011, so we instead used newly commissioned Osmos and Red4K LBNL 250 micron thick, fully depleted p-channel $4096 \times 4096$ pixel CCDs built by Ohio State University Astronomy Department. The goal was to detect the optical counterparts to binary millisecond pulsars and, when possible, obtain full phase-resolved light curves in each of the three filters. 

The eight fields observed are all discussed in \citet{Ray2012} with ephemerides provided by the radio follow-up team. Table~\ref{tab:1} gives the fields and the observations made of each of them along with their measured magnitudes in various filters or upper limits. Previously, three of these systems have been detected by \citet{Breton2013}, J0023+0923, J1810+1744,  and J2215+5135. According to their observations, only J2215+5135 and J1810+1017 are bright enough for our campaign to have been able to detect a signal, and these are, in fact, the two objects for which we obtained complete phase-resolved coverage. Additionally, J1816+4510 was identified by \citet{Kaplan2013} as being bright enough to be identified in the Digitized Sky Survey, and we confirm detection of this source. In all, we detect four counterparts: J1810+1744, J1816+4510, J2214+3000, and J2215+5135, one of which, J2214+3000 is a novel detection. The relevant details of the observations of the eight fields observed is summarized in Table~\ref{tab:1} and discussion of the objects proceeds in the following two subsections.

Conditions over the observing runs often varied with intermittent cloud cover occurring on some nights and considerable particulate matter observed in the atmosphere on the observing run of MJD 55798 to 55803. In instances of non-photometric conditions, the data was only used for difference photometry rather than absolute photometric calibration. Additional electronic noise and interference patterns affected observations on MJD 55452 as well as the entire run that used the Red4K CCDs with the latter issue being so prohibitive as to make standard reduction schema, in particular flatfielding, impossible. In both of those cases, difference photometry was only attempted if calibration of the images indicated minimal systematic uncertainties (that is, if the background counts were relatively constant in the area of interest).

The images were reduced using the \texttt{ccdproc} IRAF routine for bias subtraction and flatfielding while, if necessary, astrometric solutions were obtained using \texttt{imcoords}. If needed, as in the case of obtaining upper limits, images taken with the same instrument in an observing run were combined using the \texttt{imalign} and \texttt{imcombine} routines. Aperture photometry was done using the \texttt{apphot} package for a range of apertures with absolute photometric calibration done using large apertures and difference photometry done using small apertures (the precise sizes chosen on the basis of the minimum of the photometric errors in the case of small apertures and in consideration of field crowdedness in the case of large apertures). Absolute photometry was calibrated using interspersed observations of Landolt standard fields \citep{Landolt1992} which were used to confirm both the photometric offsets, variation with airmass, and any color-correction terms. For J1810+1744 and J2215+5135, secondary standards for difference photometry were chosen on the basis of observed photometric stability. For both objects, we measured the zero-point calibration to be precise to within bootstrap errors of 0.1 mag.

Phase-resolution was done using published ephemerides where possible and some provided by means of private communication. The orbital phase of each observation was computed by applying a heliocentric correction to the observation time ($t_{\rm obs}$), subtracting the result from $\phi = 0$ time as measured from the radio ephemeris ($T_0$) for each observation and finding the fractional remainder of the orbit. Thus 
\begin{equation}
\phi = \frac{\left(t_{\rm obs} - T_0\right) \bmod{P_{\rm orb}}}{P_{\rm orb}} \label{eq:phase}
\end{equation}
is used where, by convention, $\phi = 0$ is the phase of the epoch of the ascending node. The expected optical signal should exhibit a maximum due to irradiation at $\phi = 0.75$ while the minimum associated with the line-of-sight visibility of the nonirradiated side occurs at $\phi=0.25$. The normally smaller ellipsoidal variations exhibit peak brightness at $\phi = 0$ and $\phi = 0.5$ respectively. 

In the case of J1810+1744, \citet{Breton2013} report that a neighboring star is a possible contaminant near minimum. Our attempts to detect these faint contaminant stars by combining images near minimum and performing PSF-fitting were unsuccessful, and so we concluded that aperture photometry was likely to be accurate enough for the difference photometry analysis.

Below we discuss our findings for each object individually together with related results from previous work.

\subsection{J2214+3000}
\label{subsec:J2214}

\begin{figure}[t!!]
\vskip 0.7truecm
\begin{center}
\includegraphics[scale=0.4, angle=270]{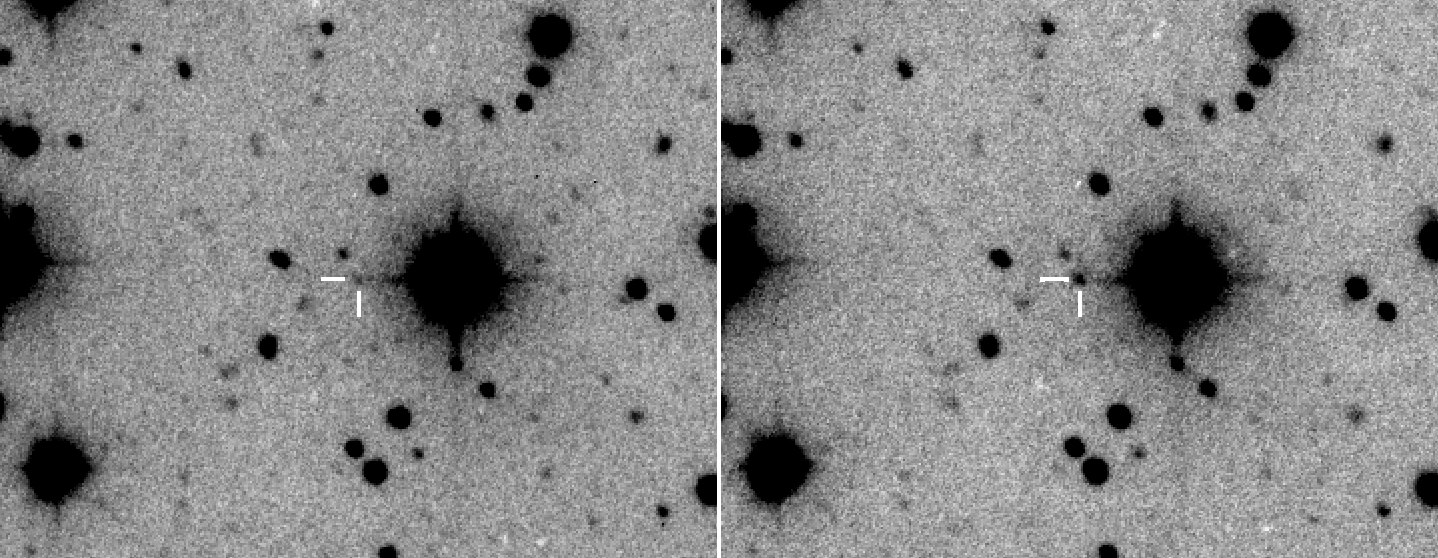}
\caption{\label{fig:1}Indicated by the white cross-hatch, the location of J2214+3000 as reported by the ephemeris of \citet{Ransom2011}. The image on the top is a composite of 10 \texttimes{} 600 s R-band images at phases $\phi = 0.701 \pm 0.118$ while the image on the bottom is a composite of 10 \texttimes{} 600 s R-band images at $\phi = 0.355 \pm 0.0818$. }
\end{center}
\end{figure}

J2214+3000 was identified by \citet{Ransom2011} and the radio data therein provided indicates a rotational period of the pulsar of 3.1 ms, a binary orbital period of 9.8 hrs, and a binary mass function $f = 8.7 \times 10^{-7} M_{\odot}$. Of particular interest to this work is that the precise position of the source (epoch J2000) is given at $22^{\rm h}\;14^{\rm m}\;38\fs8460(1)$ in right ascension and $+30\degrees\;00\amin\;38\farcs234(4)$ in declination. To within astrometric errors, this is consistent with the coordinates where we detected the companion. 

We combined R-band images at $\phi = 0.75 \pm 0.1$ to obtain the detection listed in Table~\ref{tab:1} while the lower limits of R included combinations of images from all other periods. Additionally, combinations of the B images which range from $0.1 \le \phi \le 0.6$ yielded a null result. See Figure~\ref{fig:1} for a comparison image.

The detection limits of the observing capabilities of the facilities at MDM observatory are such that we were unable to obtain a complete photometric light curve for J2214+3000. For our observing runs, we determined it was an unsuitable candidate for complete phase coverage and model-fitting in spite of its detection for three reasons: 1) its long orbital period makes obtaining full phase coverage in multiple filters prohibitive, 2) it is the faintest detection in our sample and characterization of its light curve outside of peak brightness near $\phi \sim 0.75$  would be unlikely, and 3) there is source confusion with a nearby bright $R = 14.1$ some $14''$ to the west, and the bright source's diffraction spike is oriented in exactly the same direction causing source confusion for our deepest observations which is why, for example, we can only report upper limits for minimum in Table~\ref{tab:1} in spite of having a combined exposure time > 100 minutes within $\Delta \phi = 0.1$ of $\phi = 0.25$.

\subsection{J1816+4510}

J1816+4510 was first identified as an eclipsing binary millisecond pulsar by the 350 MHz Green Bank North Celestial Cap Survey with a pulsar rotational period of 3.2 ms, binary orbital period of 8.6 hrs, and binary mass function $f = 1.7 \times 10^{-3} M_{\odot}$. Although its mass function is consistent with redbacks, its optical counterpart was identified by \citet{Kaplan2012} as being a white dwarf with R = 18.3 and $T_{\rm eff} = 15\ 000$ K with no evident phase-dependent heating, though typical effective irradiation luminosities measured in other short period binary systems would not be detectible with such an intrinsically hot companion. Our six observations made of this object are consistent with no variability to within measured uncertainties, so no constraints can be placed on system parameters. Determination of such will require a more-dedicated observational campaign similar to the ones we conducted on J1810+1744 or J2215+5135.

\subsection{J1810+1744}

J1810+1744 is a $P_{\rm rot} = 1.7$ ms pulsar discovered with the 350 MHz frequency channel at the Green Bank Telescope (GBT) \citep{Hessels2011} with an orbital period of $P_{b} = 3.6$ hrs and binary mass function $f = 4.4 \times 10^{-5} M_{\odot}$. Our observations were able to isolate the optical signature of the secondary over the course of a number of runs, and complete phase coverage in three bands was obtained as reported in Table~\ref{tab:1}. At minimum, the source is near the detection limits, and for approximately half the observations taken between phases $0.15 \le \phi \le 0.35$ we are only able to reliably report upper limits. The measurements along with the uncertainties are plotted in Figure~\ref{fig:2}.

Our measurements are somewhat in tension with the reported light curve fits of \citet{Breton2013}. Using the \citet{Lupton2005} transformation equations, our R and V light curves and the $i$ and $g$ light curves from \citet{Breton2013} imply the same r-band light curve to within $|\Delta r| < 0.3$. However, our B light curve when matched to their $g$ light curve gives a $r$-band light curve that differs at minimum by more than $|\Delta r| > 3$. We propose therefore that there is a much steeper drop in the blue-end of the spectrum for this object which would imply that their best-fit $g$ light curve is too bright at minimum. To provide a level of corroboration, we note that a single faint $g = 22.8$ detection near minimum ($\phi = 0.2$) \emph{is} consistent with our observations in the sense that if we take that measurement to be the $g$ magnitude at minimum and redo the transformation at that phase, we obtain a a prediction for an $r$-band magnitude that is within $|\Delta r| < 0.3$ of the initial prediction. 

We report on how we obtained a set of best-fit model light curves on the basis of these observations in Section~\ref{sec:J1810+1744 Model Fits}.

\subsection{J2215+5135}

 J2215+5135 was also discovered with the 350 MHz GBT search with a $P_{\rm rot} = 2.6$ ms, $P_{b} = 4.1$ hrs, a binary mass function $f = 3.6 \times 10^{-3} M_{\odot}$, and observed radio eclipses. Full phase coverage was obtained as reported in Table~\ref{tab:1} and the object's photometry is well-constrained at minimum. As a redback, this object exhibits less orbital modulation than black widows consistent with a larger-mass companion.

\citet{Breton2013} fits four observations in the $i$-band to a light curve that, when transformed using the equations of \citet{Lupton2005} to our R-band light curve, give a prediction of $r$ at $\phi = 0.45$ consistent to within $|\Delta r| < 0.1$ of the prediction when transforming their observation of $g = 19.2$ to our observations of B, V, and R at the same phase. 

We report on how we obtained our best-fit model light curves on the basis of these observations in Section~\ref{sec:J2215+5135 Model Fits}.

\subsection{J0023+0923}

J0023+0923 was identified with the 350 MHz GBT search with a $P_{\rm rot} = 3.1$ ms, $P_{b} = 3.4$ hrs, a binary mass function $f = 2.4 \times 10^{-6} M_{\odot}$ but with no observed radio eclipses. The optical counterpart was detected by \citet{Breton2013} with a brightness at maximum of $i = 21.7$ according to their best-fit model. We report null results in B, V, and R-bands in Table~\ref{tab:1}.

\subsection{J1745+1017}

J1745+1017 was discovered with the Effelsberg Radio Telescope using the 1.32 GHz channel and has $P_{\rm rot} = 2.7$ ms, $P_{b} = 17.5$ hrs, a binary mass function $f = 1.4 \times 10^{-6} M_{\odot}$ \citep{Barr2013} Performing a photometric analysis at the reported position of $17^{\rm h}\;45{\rm m}\;33\fs8371(7)$ $+10\degrees\;17\amin\;52\farcs523(2)$ of images combined within a few hours of $\phi = 0.75$  yielded the null result as reported in Table~\ref{tab:1}.

\subsection{J2047+1053}

According to \citet{Ray2012}, J2047+1053 was discovered at the Nan\c{c}ay Radio Telescope and has $P_{\rm rot} = 4.3$ ms, $P_{b} = 2.9$ hrs, a binary mass function $f = 2.3 \times 10^{-5} M_{\odot}$. We report null results in B, V, and R-bands in Table~\ref{tab:1}.

\subsection{J2234+0944}
According to \citet{Ray2012}, J2234+0944 was discovered using the Parkes Radio Telescope and has $P_{\rm rot} = 3.6$ ms, $P_{b} = 10$ hrs, a binary mass function $f = 1.7 \times 10^{-6} M_{\odot}$. We report null results in the R-band in Table~\ref{tab:1}.

\begin{table*}
\caption{\label{tab:1}}
\begin{center}
\begin{tabular}{cccccccc}
Object & MJD & Exposure & Filters & Airmass & Telescope & Instrument & Magnitude\tabularnewline
\hline 
\multirow{3}{*}{J0023+0923} & 55803 & 42 \texttimes{} 300 s & R & < 1.33 & McGraw-Hill & Red4K & > 23.52\tabularnewline
 & 55801 & 5 \texttimes{} 300 s & V & < 1.4 & Hiltner & Osmos & > 23.23\tabularnewline
 & 55802 & 40 \texttimes{} 300 s & B & < 1.21 & Hiltner & Osmos & > 24.21\tabularnewline
\hline 
\multirow{3}{*}{J1745+1017} & 55329 \textpm{} $ $1 & 100 \texttimes{} 300 s & R & < 3.11 & Hiltner & Echelle & \multirow{3}{*}{> 24.78}\tabularnewline
 & 55353 \textpm{} 1 & 125 \texttimes{} 300 s & R & < 2.3 & McGraw-Hill & Nellie & \tabularnewline
 & 55365 & 20 \texttimes{} 600 s & R & < 1.607 & McGraw-Hill & Nellie & \tabularnewline
\hline 
\multirow{9}{*}{J1810+1744} & 55329 & 42 \texttimes{} 300 s & R & < 2.084 & Hiltner & Echelle & \multirow{5}{*}{(22.74, 19.63, 20.31)}\tabularnewline
 & 55362 \textpm{} 1 & 80 \texttimes{} 300 s & R & < 1.44 & McGraw-Hill & Templeton & \tabularnewline
 & 55356 & 9 \texttimes{} 300 s & R & < 1.285 & Hiltner & Nellie & \tabularnewline
 & 55448 & 64 \texttimes{} 100 s & R & < 2.2 & Hiltner & Echelle & \tabularnewline
 & 55800 & 6 \texttimes{} 300 s & R & < 1.04 & Hiltner & Osmos & \tabularnewline
 & 55714 & 30 \texttimes{} 120 s & V & < 1.248 & Hiltner & Templeton & \multirow{2}{*}{(23.57, 19.82, 20.52)}\tabularnewline
 & 55715 & 25 \texttimes{} 600 s & V & < 1.413 & Hiltner & Templeton & \tabularnewline
 & 55715 & 40 \texttimes{} 300 s & B & < 1.58 & Hiltner & Templeton & \multirow{2}{*}{(24.11, 19.93, 20.68)}\tabularnewline
 & 55797 & 37 \texttimes{} 300 s & B & < 2.64 & Hiltner & Osmos & \tabularnewline
\hline 
J1816+4510 & 55743 & 6 \texttimes{} 300 s & R & < 1.079 & McGraw-Hill & Templeton & 18.27\tabularnewline
\hline 
\multirow{3}{*}{J2047+1053} & 55802 & 40 \texttimes{} 300 s & B & < 1.204 & Hiltner & Osmos & > 24.20\tabularnewline
 & 55744 & 20 \texttimes{} 300 s & R & < 1.16 & McGraw-Hill & Templeton & \multirow{2}{*}{> 22.92}\tabularnewline
 & 55802 & 36 \texttimes{} 300 s & R & < 1.3 & McGraw-Hill & Red4K & \tabularnewline
\hline 
\multirow{6}{*}{J2214+3000} & 55331 & 2 \texttimes{} 300 s & R & 1.233 & Hiltner & Echelle & \multirow{5}{*}{(> 23.42, 22.64)}\tabularnewline
 & 55363 & 10 \texttimes{} 600 s & R & < 1.305 & McGraw-Hill & Templeton & \tabularnewline
 & 55356 & 8 \texttimes{} 300 s & R & < 1.57 & Hiltner & Nellie & \tabularnewline
 & 55449 & 39 \texttimes{} 600 s & R & < 1.61 & Hiltner & Echelle & \tabularnewline
 & 55447 & 220 \texttimes{} 60 s & R & < 2.214 & Hiltner & Echelle & \tabularnewline
 & 55803 & 57\texttimes{} 300 s & B & < 2.08 & Hiltner & Osmos & > 24.75\tabularnewline
\hline 
\multirow{8}{*}{J2215+5135} & 55356 & 3 \texttimes{} 300 s & R & < 1.6 & Hiltner & Nellie & \multirow{4}{*}{(19.70, 18.53, 18.94)}\tabularnewline
 & 55452 & 29 \texttimes{} 600 s & R & < 2.31 & Hiltner & Echelle & \tabularnewline
 & 55743 \textpm{} $ $2 & 55 \texttimes{} 300 s & R & < 1.819 & McGraw-Hill & Templeton & \tabularnewline
 & 55743 & 75 \texttimes{} 100 s & R & < 1.3 & McGraw-Hill & Templeton & \tabularnewline
 & 55747 & 23 \texttimes{} 200 s & V & <1.25 & McGraw-Hill & Templeton & \multirow{2}{*}{(20.19, 18.73, 19.27)}\tabularnewline
 & 55800 & 51 \texttimes{} 300 s & V & < 1.3 & Hiltner & Osmos & \tabularnewline
 & 55798 \textpm{} $ $2 & 96 \texttimes{} 300 s & B & < 1.69 & Hiltner & Osmos & \multirow{2}{*}{(20.77, 18.92, 19.62)}\tabularnewline
 & 55716 & 11 \texttimes{} 300 s & B & < 1.189 & Hiltner & Templeton & \tabularnewline
\hline 
J2234+0944 & 55449 & 10 \texttimes{} 300 s & R & < 1.957 & Hiltner & Echelle & > 22.69\tabularnewline
\hline 
\end{tabular}
\vskip 0.2truecm
Table of observations included in this paper. Telescopes
indicated are either the 2.4 m Hiltner Telescope or the 1.3 m McGraw-Hill
telescope. Instrument specifications are listed in Section~\ref{sec:Observations and Data Analysis}. The magnitude limits are either given as the maximum image-combined
3-$\sigma$ upper limits, image-combined detections near $\phi=0.75$,
or (maximum, minimum, and quadrature) magnitude model fits for the irradiated
light curves measured for J1810+1744 and J2215+5135. In
the case of J1810+1744, approximately half the observations within $0.15 \leq \phi \leq 0.35$ yielded only upper limits. Since our model fits are also worst at minimum flux, we note that the true maximum magnitude could be different by as much as $\Delta {\rm mag} = 1$, as can be seen in Figure~\ref{fig:2}. In
the case of J2214+3000, detections at maximum in R are reported as
well as an upper-limit for the minimum null detection in both R and
B, which is treated with skepticism due to source confusion as described in Section~\ref{subsec:J2214}. 
\vskip 0.2truecm
\end{center}
\end{table*}

\section{Modeling and Parameter Estimation}
\label{sec:Modeling and Parameter Estimation}

\begin{figure}[t!!]
\vskip 0.7truecm

\caption{\label{fig:2}The upper part of each plot includes the phase-resolved light curves in B, V, and R filters for J1810+1744 (top) and J2215+5135 (bottom). Vertical error bars are the photometric uncertainties while the horizontal error bars are the exposure time for each observation. $3-\sigma$ upper limits for null detections are indicated by downward pointing arrows. Best-fit PHOENIX models as reported in Table~\ref{tab:2} are indicated by the solid lines where $\epsilon = 10$ was the efficiency factor chosen for the model fit of J1810+1744. The lower part of each plot shows the magnitude residuals from the best-fit model.}

\begin{tabular}{c}
\includegraphics[scale=0.6, trim=90 20 0 0]{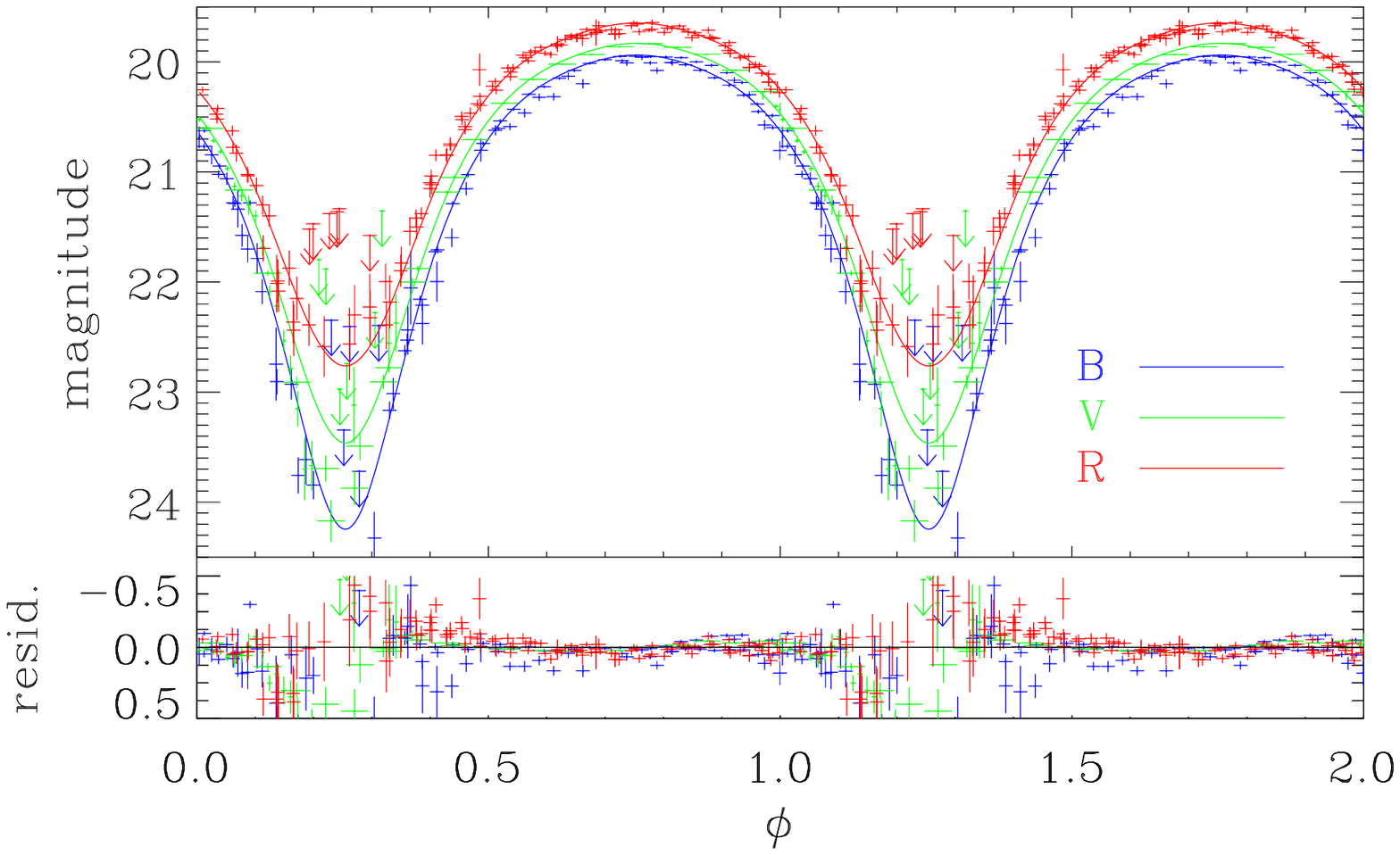}
\tabularnewline
\includegraphics[scale=0.6, trim=90 20 0 0]{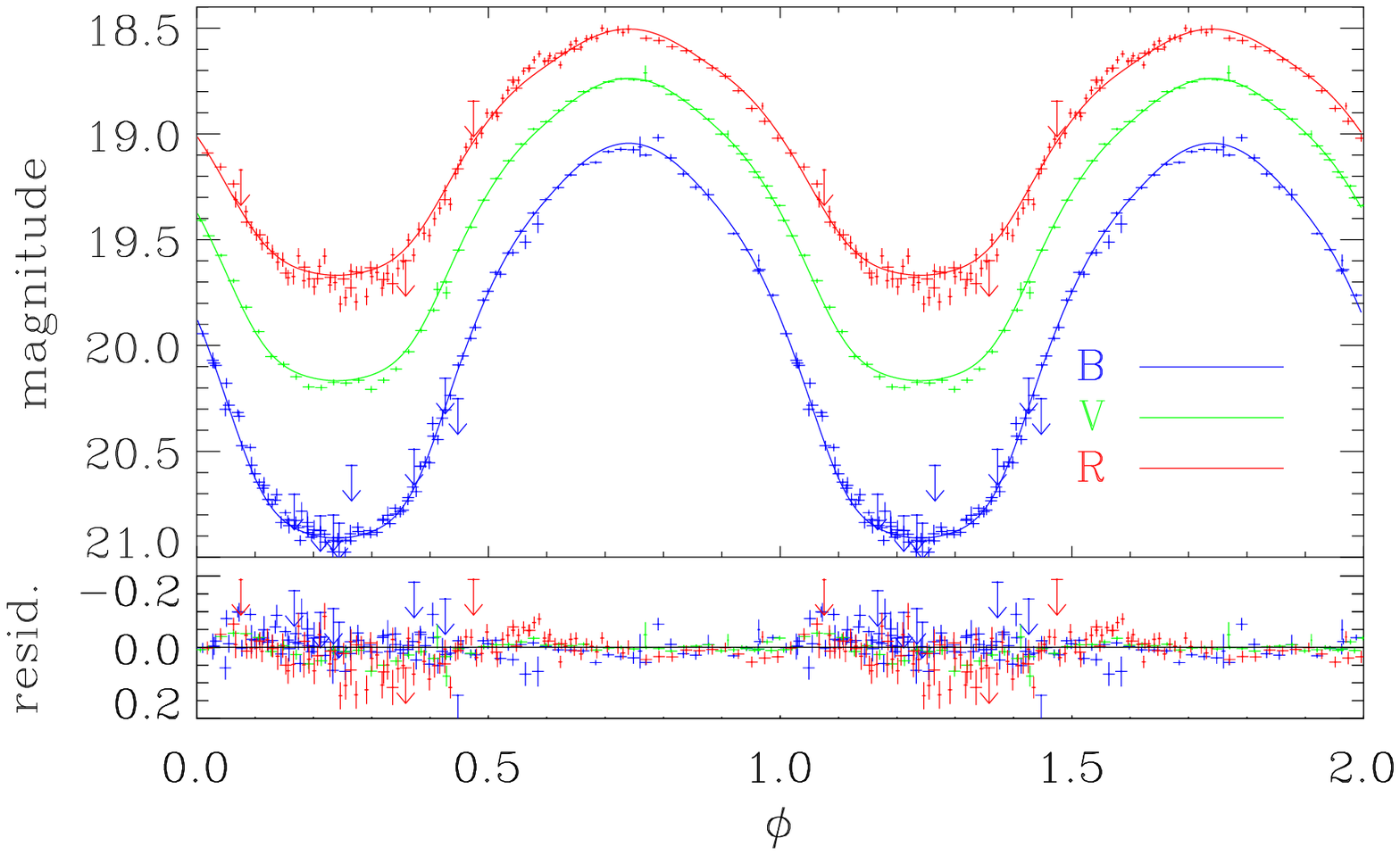}
\tabularnewline 
\end{tabular}

\end{figure}

We show in Figure~\ref{fig:2} the full, phase-matched BVR light curves for the optical counterparts of J1810+1744 and J2215+5135 respectively. The magnitude errors and phase-scaled observing times are indicated by the vertical and horizontal error bars respectively. Near minimum under circumstances involving cloud cover or in cases where the integrated flux over the observation was below the detection threshold for the telescope and/or the instrument, upper limits are reported. The solid curves in each Figure are the best-fit model light curves obtained by means we now describe.

To model our light curves, we used the ELC code of \citet{Orosz2000} which tracks the visibility of the Roche lobe geometry of the system and invokes a model atmosphere with the appropriate gravitational and limb darkening at each grid point. We used the code in its ``millisecond pulsar mode'' which models a circularized, tidally locked system with $P_{\rm orb}$ and $x$ specified from the radio ephemeris data as reported in Table~\ref{tab:2}. The ELC code allows for a point-source illumination coming from the position of the neutron star and treats the reprocessed heating of the companion as a single-iteration reflection effect. 

Model atmospheres of a large range of effective temperatures are required to constrain the companion profiles of these systems since the underlying star can be a cool dwarf while the irradiated side can exhibit temperatures that exceed $> 10\ 000\ {\rm K}$. We use both the NextGen \citep{NextGen} and the newly tabulated PHOENIX atmospheres \citep{PHOENIX} to model the stellar temperatures below $< 10\ 000\ {\rm K}$ while the ATLAS9 atmospheres \citep{ATLAS9} are used to model grid cells that are at higher temperatures.

The remaining free parameters are the following:

\begin{itemize}
\item \emph{Inclination angle $(i)$.} There are two  effects seen as the inclination angle of the system increases. The largest effect is that the line-of-sight visibility to the illuminated and nonilluminated side of the companion is affected greatly. Drops in flux around the $\phi = 0.25$ phase become more significant with higher inclination angles so that an angle of $i = 90^{\circ}$ gives the greatest variation from maximum to minimum light while an angle of $i = 0^{\circ}$ gives no variation at all. A smaller but still measurable effect in both light curves comes from ellipsoidal variations that are also stronger at the highest inclination angles. According to Roche Lobe geometry in such tidally-locked systems, light curve ellipsoidal variations have flux peaks at phases $\phi = 0$ and $\phi = 0.5$ while their minima are at $\phi = 0.25$ and $\phi = 0.75$.

\item \emph{Mass ratio $(Q = M_{\rm NS}/M_{\rm c})$}. A prior constraint of possible mass ratios can be determined by assuming that neutron stars conservatively must be between $1 M_\Sun \le M_{\rm NS} \le 3.6 M_\Sun$, though masses at both the high end and low end are very unlikely \citep{Baumgarte2000,Strobel2001}. As a higher $Q$ implies a larger system separation to accommodate the observed orbital period, this will necessarily imply a larger inclination angle as the projected semi-major axis on the sky is fixed. To accommodate the same swing from dayside to nightside flux, the increased $Q$ also implies a larger irradiation flux. Owing to the degeneracies between these parameters, radial velocity measurements of the companion through spectroscopic campaigns have been the usual way of constraining $Q$ as in \citet{vanKerkwijk2011} or \citet{Romani2012}. However, in the case of J2215+5135, the light curve is measured well enough to give model constraints on $Q$ and $i$ that only permit a narrow range of neutron star masses as discussed in Section~\ref{sec:J2215+5135 Model Fits}.

\item \emph{Effective temperature $(T_{\rm eff})$}. The underlying atmospheric structure of the star is parameterized by a single intensity-weighted mean effective temperature that, absent an irradiative heating effect, constrains the observed magnitude variations to first order.  Color constraints can be used to obtain distance estimates, but we look at such constraints only after the model fits were made owing to degeneracies with reddening effects. (See Section~\ref{sec:J2215+5135 Model Fits}.) The overall temperature structure of the star is significantly altered when an irradiation heat source is present meaning that the effective temperature of the star as given by the observationally-based blackbody calculation $T = (L \sigma^{-1} A^{-1})^{1/4}$ is higher than the model effective temperature. The primary difference between the PHOENIX and NextGen models are that best-fit PHOENIX models occur at approximately $\Delta T_{\rm eff} \sim 200 \ {\rm K}$ hotter than the NextGen models.

\item \emph{Irradiation efficiency ($\epsilon$}). Owing to fact that ELC code was developed to model x-ray binary systems, this parameter is realized in ELC as a combination of ``x-ray luminosity'' ($L_x$) modeled to emanate from a point source located at the position of the neutron star and the bolometric albedo of the reflection effect $a$ that is dependent upon the response of a stellar atmosphere to an irradiation source. Typically, convective atmospheres yield a reflection effect albedo of $a \sim 0.5$ while radiative atmospheres have an albedo $a = 1.0$ \citep{Rucinski1969}. Since this albedo is due entirely to the conditions of the envelope (convective envelopes are constrained by the condition of adiabatic temperature gradients while radiative envelopes are constrained by radiative flux equilibrium), the ELC code cannot distinguish between different $a L_x =\ {\rm constant}$ models without more information as to the incident flux or the detailed envelope structure. The only meaningful observable constraints applicable then is a comparison of the irradiation necessary with the spin-down luminosity of the pulsar, which we give as the irradiation efficiency
\begin{equation}
\epsilon = \frac{aL_x}{L_{\rm SD}} =aL_x\frac{P_{\rm rot}^{3}}{ 4\pi^{2}I\dot{P}_{\rm rot}}.
\end{equation} The assumption that $\epsilon \le 1$ is shown to be problematic for J1810+1744 as we discuss in Section~\ref{sec:J1810+1744 Model Fits}.

\item \emph{Roche lobe filling fraction $(f_{\mathrm{Roche}})$}. A constraint on the size of the secondary, this parameter affects ellipsoidal variations and the $\log g$-dependent gravity darkening effects for a known mass ratio $(Q)$. Indeed, for a given $i$ and $Q$, $f_{\rm Roche}$ and $\epsilon$ are both strongly constrained by the particular shape of the observed light curve at its brightest due to a competition between the heating effects that are flux-enhanced at $\phi = 0.75$ and the ellipsoidal variations that are flux-suppressed at $\phi = 0.75$. In highly irradiated cases, observations at minimum are needed to be able to constrain the filling factor as the relative signal from the ellipsoidal variations becomes more difficult to detect on a dayside increasingly dominated by irradiated flux. Other black widow systems are measured to be very close to filling their Roche lobes, \citep{Reynolds2007, Romani2012, Breton2013} consistent with the wind-driven outflows generally considered necessary for radio eclipses \citep{Eichler1995} and binary evolution scenarios \citep{Chen2013}. Additionally, transitioning redbacks such as J1023+0038 and J18245-2452 must satisfy the Roche-lobe overflow condition $f_{\mathrm{Roche}} = 1$ in order for accretion to happen.

\item \emph{Phase shift $(\Delta \phi)$}. While the orbital parameters including the phase-timing are set by the radio data, it is possible that the heating of the secondary is offset by a certain amount due, for example, to an angular offset between the orbital relationship between the stars and the location of the stand-off shock.

\end{itemize}
\begin{table*}
\begin{center}
\caption{\label{tab:2}}
\begin{tabular}{c|cc|cc}
\multicolumn{5}{c}{\emph{Parameters Observed and Derived from Radio Observations}}\tabularnewline
\multicolumn{5}{c}{}\tabularnewline
parameters & \multicolumn{2}{c|}{J1810+1744} & \multicolumn{2}{c}{J2215+5135}\tabularnewline
\hline
$P_{\rm rot}\ {\rm (ms)}$ & \multicolumn{2}{c|}{1.66} & \multicolumn{2}{c}{2.61}\tabularnewline
$\dot{P}_{\rm rot}\ (10^{-20})$ & \multicolumn{2}{c|}{0.46} & \multicolumn{2}{c}{2.34} \tabularnewline
$DM\ ({\rm pc\ cm^{-3}})$ & \multicolumn{2}{c|}{39.66} & \multicolumn{2}{c}{69.19}\tabularnewline
$P_{\rm orb}\ {\rm (h)}$ & \multicolumn{2}{c|}{3.56} & \multicolumn{2}{c}{4.14}\tabularnewline
$T_0\ ({\rm MJD})$ & \multicolumn{2}{c|}{55130.048136022} & \multicolumn{2}{c}{55186.164695228}\tabularnewline
$x\ ({\rm lightsec})$ & \multicolumn{2}{c|}{0.095} & \multicolumn{2}{c}{0.468}\tabularnewline
\hline
\hline
\multicolumn{5}{c}{}\tabularnewline
\multicolumn{5}{c}{\emph{ELC Model Parameters}}\tabularnewline
\multicolumn{5}{c}{}\tabularnewline
 & NextGen & PHOENIX & NextGen & PHOENIX\tabularnewline
\hline 
$i\ (^\circ)$ & $56.75\pm2.25$ & $54.75\pm2.75$ & $51.7_{-1.5}^{+2.3}$ & $51.6_{-2.1}^{+2.7}$\tabularnewline
$Q$ & $30 \pm 7$ & $29.5 \pm 6.5$&$5.7_{-0.15}^{+0.3}$ & $6.2\pm0.25$\tabularnewline
$T_{{\rm eff}}\ ({\rm K})$ & $4525\pm175$ & $4425\pm225$ & $3925\pm20$ & $3790_{-25}^{+35}$\tabularnewline
$\epsilon$ &$10.4\pm5.4$ &$7.9\pm4.7$ & \multicolumn{2}{c}{$0.083\pm0.001$}\tabularnewline
$f_{\rm Roche}$ & \multicolumn{2}{c|}{$1.000\pm0.007$} & \multicolumn{2}{c}{$1.000\pm0.008$}\tabularnewline
$\Delta\phi$ & \multicolumn{2}{c|}{$4.8\times10^{-3}\pm9\times10^{-4}$} & \multicolumn{2}{c}{$-9.5\times10^{-3}\pm5\times10^{-4}$}\tabularnewline
\hline 
\hline 
\multicolumn{5}{c}{}\tabularnewline
\multicolumn{5}{c}{\emph{Parameters Derived from Model Fits}}\tabularnewline
\multicolumn{5}{c}{}\tabularnewline
 & NextGen & PHOENIX & NextGen & PHOENIX\tabularnewline
\hline 
$M_{{\rm NS}}\ (M_\Sun)$ & \multicolumn{2}{c|}{$1.0$ to $3.6$} & $1.97_{-0.05}^{+0.08}$ & $2.45_{-0.11}^{+0.22}$\tabularnewline
$M_{{\rm c}}\ (M_\Sun)$ & $0.0710\pm0.0273$ & $0.0735\pm0.0285$ & $ $$0.345_{-0.007}^{+0.008}$ & $0.396\pm0.045$\tabularnewline
$T_{\rm hot}\ ({\rm K})$ & $14\ 500\pm1000$ & $13\ 500\pm1100$ &$5073_{-26}^{+17}$ &  $4899_{-23}^{+34}$\tabularnewline
$K\ ({\rm km\ s^{-1}})$& $421\pm98$ &$414\pm91$&$338_{-9}^{+17}$ &$367\pm15$ \tabularnewline 
\end{tabular}
\vskip 0.2truecm
Table of system parameters for J1810+1744 and J2215+5135 obtained from the radio data, best-fit to the ELC model using either NextGen or PHOENIX atmospheric models, and derived on the basis of the fits. Reported uncertainties were found by computing the $\chi^2$ statistic and finding the $\Delta \chi^2 = 1$ ranges for each individual parameter. In the case of $\Delta \phi$, the NextGen and PHOENIX modeling gives the same fit because these parameters are affected only by the orientation with respect to line of sight rather than the specific Roche lobe geometry or irradiation environment. Additionally, for J2215+5153, the $\epsilon$ and $f_{\rm Roche}$ values exhibit the same constraints regardless of the model atmosphere used. In the case of J1810+1744, $f_{\rm Roche}$ was similarly constrained, but since no best-fit was possible, the reported range for neutron star masses was assumed as a prior. $\Delta \chi^2 = 1$ limits on $i$, $Q$, $T_{\rm eff}$ were then found for each $\epsilon$-value and collated to give the limits presented here. The range of the other three derived parameters were collated on a similar basis. Constraints on $\epsilon$ values are shown as to which values permit $\chi^2$ fits in assumed mass range.
\vskip 0.2truecm
\end{center}
\end{table*}

In addition to the above fitted parameters, derived parameters of interest can be calculated for each model including the masses of each component using Equation~\ref{eq:mass}, the dayside temperature which, assuming a blackbody, follows the equation 
\begin{equation}
T_{\rm hot} = \left(T_{\rm eff}^4 + \frac{\epsilon L_{\rm SD} \sin ^2 i}{4 \pi \left(x (1+Q)\right)^2 \sigma}\right)^{1/4}
\end{equation}
and the predicted semi-amplitude of the radial velocity curve given by
\begin{equation}
K = 212.9 \sin i \left(\frac{M_{\rm NS}}{P_{\rm orb} (1 + 1/Q)^2}\right)^\frac{1}{3}\ {\rm km\ s^{-1}}.
\end{equation}

For both objects, we searched for and obtained best-fit models using a variety of methods including a genetic algorithm search and a grid search based on the $\chi^2$ statistic. We were successful in finding a best-fit $\chi^2$ minimum when varying all six parameters for J2215+5135 for which $\Delta \chi^2$ confidence intervals were then be calculated. On this basis, the best-fit parameters are tabulated in Table~\ref{tab:2}. Best-fit values for J1810+1744 could only be ascertained for $\Delta \phi$ and $f_{\rm Roche}$ while limits on $i$, $Q$, and $T_{\rm eff}$ could only be placed on the basis of assumed prior probabilities for $M_{\rm NS}$ and $\epsilon$. Indeed, the global best-fit models for J1810+1744 were beyond the limits of physically plausible neutron star masses and amounts of irradiation. We discuss this issue more in Section~\ref{sec:J1810+1744 Model Fits} while Table~\ref{tab:2} lists the either the best-fit model parameters and confidence intervals based on $\Delta \chi^2 = 1$ or, for the parameters of J1810+1744 where no best-fit was possible, the range of likely values given the condition that $1.0 M_\Sun \le M_{\rm NS} \le 3.6 M_\Sun$.

\subsection{J1810+1744 Model Fits}
\label{sec:J1810+1744 Model Fits}

\begin{figure}[t!!]
\vskip 0.7truecm

\caption{\label{fig:3} J1810+1744 best-fit models using the PHOENIX model atmospheres (top) and the NextGen model atmospheres (bottom) in the $Q$ vs. $i$. The solid green line traces the location of the best-fit models for efficiencies in order from bottom to top of the plot as $\epsilon = \left(1.0, 1.3, 1.6, 2.0,2.5,3.2,4.0,5.0,6.3,7.9,10.0,12.6,15.8,20,25.1,31.6 \right)$ with the points indicated with crosses. Black, dark gray, and light gray respectively correspond to the $\Delta \chi^2 = 1, 4,\ {\rm and}\ 9$ contours projected for efficiencies of $\epsilon = 1,4.0,10.0$ for the PHOENIX models and $\epsilon = 1,5.0,12.6$ for the NextGen models. Additionally, evenly-spaced lines of constant $M_{\rm NS}$ are shown in dotted-blue while lines of constant $M_{\rm c}$ are shown in dashed-red with certain associated numerical values labeled. Note that the $\Delta \chi^2 = 1$ projection of these regions onto $Q$, $i$, $M_{\rm NS}$, and $M_{\rm c}$ are the uncertainties quoted in Table~\ref{tab:2}.}

\begin{tabular}{c}
\includegraphics[scale=0.45]{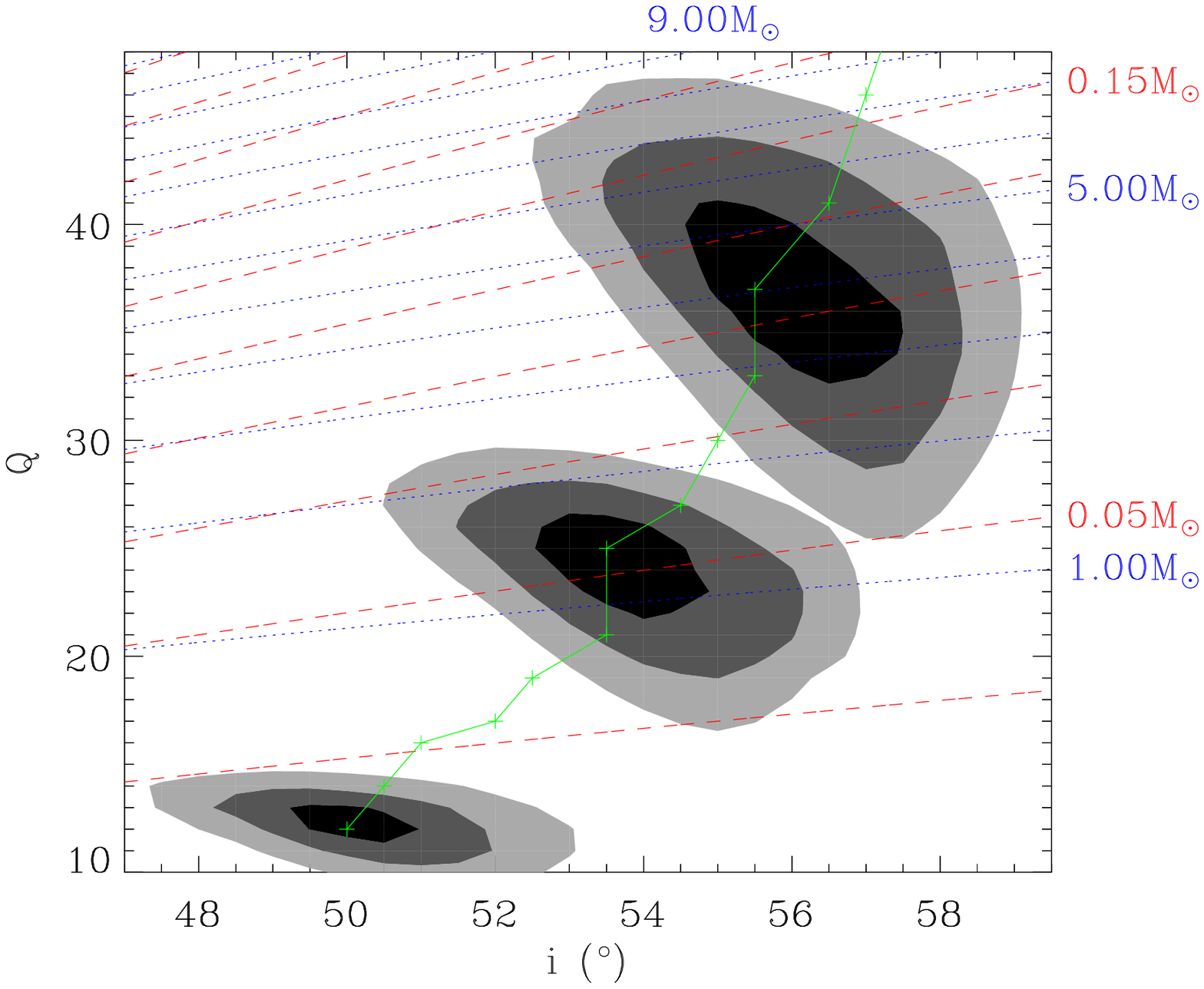}\tabularnewline
\includegraphics[scale=0.45]{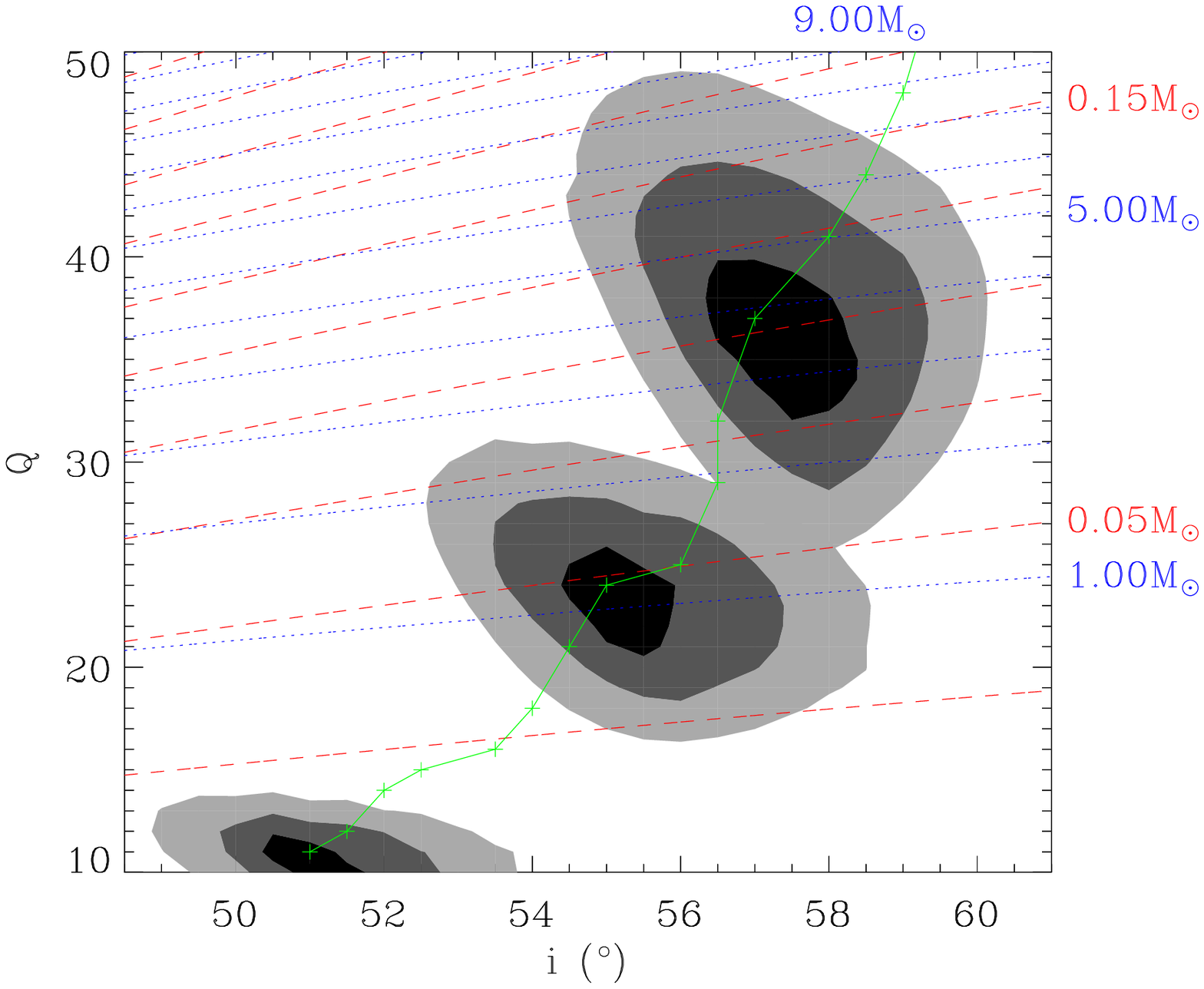}\tabularnewline 
\end{tabular}

\end{figure}

Without relaxing either the condition of $\epsilon \le 1.0$ or $M_{\rm NS} > 1.0 M_\Sun$, neither of the NextGen nor the PHOENIX models are able to reproduce the observations of the B and V bands at minimum $\phi = 0.25$. If we impose a hard-limit of $\epsilon < 1$ on our fitting routine, we find all physically plausible neutron star masses $M_{\rm NS} > 1 M_\Sun$ excluded at projected $3-\sigma$ ($\Delta \chi^2 < 9$). Allowing the efficiency to float results in the best-fit model with the lowest reduced $\chi^2/\nu = 1.7$  at $\epsilon \sim 100$  and similarly implausible neutron star masses of $M_{\rm NS} > 10 M_\Sun$. We therefore propose that 1) it is likely that $\epsilon > 1$, and 2) in spite of $\chi^2$ values favoring higher efficiencies as $\Delta \chi^2 / \Delta \epsilon < -9$, the physical implausibility of higher $\epsilon$ values means that we cannot use our fitting routine alone to distinguish between different efficiencies.

However, other system parameters are fit to a high degree of confidence as demonstrated in Table~\ref{tab:2}. For example, we find that the companion to J1810+1744 is filling its Roche Lobe to within a degree of uncertainty of $\Delta f_{\rm Roche}/f_{\rm Roche} < 1 \times 10^{-3}$. This fit applies regardless of which efficiency we choose on the basis of the shape of the light curve near minimum since its precise slope is dependent on matching the visibility to the hot side of the companion with a steepened drop in flux due to the minimum in the ellipsoidal variations. Essentially, without allowing for a nearly filled Roche lobe, the light curve's precipitous drop at $\phi = 0.25$ is too steep for any point illuminated sphere. As \citet{Romani2012} points out in their fit for a ``flyweight'' mass companion to a millisecond pulsar, allowing for a smaller-in-extent hot-spot on the illuminated side could potentially resolve tensions in the model fits, but rather than add this additional set of free parameters we use the simplest irradiation model and leave improvements in modeling the possible temperature structure for future investigations. 

A phase shift corresponding to a 62 second delay from the measured radio-ephemeris phase calculated by means of Equation~\ref{eq:phase} is robustly measured and we discuss the implications of this particular result in more detail in Section~\ref{sec:PhaseShifts}.

Because our best-fit model lies outside the range of plausible neutron star masses, we report limits in Table~\ref{tab:2} that reflect trials taken at varying values of $\epsilon = $ 1.0, 1.3, 1.6, 2.0, 2.5, 3.2, 4.0, 5.0, 6.3, 7.9, 10.0, 12.6, 15.8, 20, 25.1, and 31.6. For each efficiency, a best-fit minimum $\chi^2$ in $i$, $Q$, and $T_{\rm eff}$ is ascertained, the $i$- and $Q$-locations of which are plotted in green in Figure~\ref{fig:3}. Because the $\chi^2$ fits strongly favor higher efficiencies, simply imposing a prior constraint on the neutron star mass would return the highest neutron star mass considered. We instead calculate the areas of parameter space where, for each slice in $\epsilon$, $\Delta \chi^2 \le 1$ overlaps with areas of parameter space corresponding to $1 M_\Sun \le M_{\rm NS} \le 3.6 M_\Sun$. This gives limits on the possible values of $\epsilon$. Additionally, limits on $i$, $Q$, and $T_{\rm eff}$ are given as the maximum extent that $\Delta \chi^2 =1$ around specific values of $\epsilon$ give with the additional neutron mass constraints. To illustrate the character of this peculiar likelihood function, we have also plotted in Figure~\ref{fig:3} certain representative $\Delta \chi^2$ confidence intervals around three arbitrarily chosen values of $\epsilon$ for both the NextGen and PHOENIX models. While limits we place on system parameters overlap between the NextGen and PHOENIX atmospheres, we separately report ranges of $i$, $Q$, $T_{\rm eff}$, and $\epsilon$ for the two models in Table~\ref{tab:2} to illustrate how the model fits change under the influence of these different sets of assumptions. 

Assuming no reddening, the $B-V= 0.158$ color at maximum reported in Table~\ref{tab:1} corresponds to a color temperature lower bound of $T_{\rm color} > 8000\ {\rm K}$  \citep{Flower1996} consistent with the reported blackbody results of \citet{Breton2013}. Correcting for reddening in this direction and taking consideration of the systematic uncertainty in the zero-point of the magnitude scale yields a $B-V = 0.02 \pm 0.1$ that is consistent with $T_{\rm color} = 9600\ \pm 1500{\rm K}$ for solar metallicities. We note that because the integrated profile of the companion varies strongly with longitude, the color temperature should be cooler than the maximum $T_{\rm hot} \approx 14\ 000\ {\rm K}$ reported Table~\ref{tab:2} owing the the visibility of cooler and redder locations on the surface even at maximum.

\citet{Breton2013} describe model fits for J1810+1744 as indicating an inclination angle of $i = 48^{\circ} \pm 7^{\circ} $ and a blackbody temperature profile ranging from 4600 to 8000 K. They reject numerical light curve modeling in part due to their best-fit model yielding a physically implausible efficiency of $\epsilon = L_{x/}L_{\rm SD} = 1.5$ which is lower than the efficiencies that we find are best fit. Their lower inclination angle also accommodates a greater range of possible $f_{\rm Roche}$, though their results were consistent with $f_{\rm Roche}=1$.

\subsection{J2215+5135 Model Fits}
\label{sec:J2215+5135 Model Fits}

\begin{figure}[t!!]
\vskip 0.7truecm

\caption{\label{fig:4} J2215+5135 confidence regions around the best-fit using the PHOENIX model atmospheres (top) and the NextGen model atmospheres (bottom). Black, dark gray, and light gray  respectively correspond to the $\Delta \chi^2 = 1, 4,\ {\rm and}\ 9$ contours projected into the $Q$ vs. $i$ plane. Additionally, evenly-spaced lines of constant $M_{\rm NS}$ are shown in dotted-blue while lines of constant $M_{\rm c}$ are shown in dashed-red with certain associated numerical values labeled. Note that the $\Delta \chi^2 = 1$ projection of these regions onto $Q$, $i$, $M_{\rm NS}$, and $M_{\rm c}$ are the uncertainties quoted in Table~\ref{tab:2}.}

\begin{tabular}{c}
\includegraphics[scale=0.45]{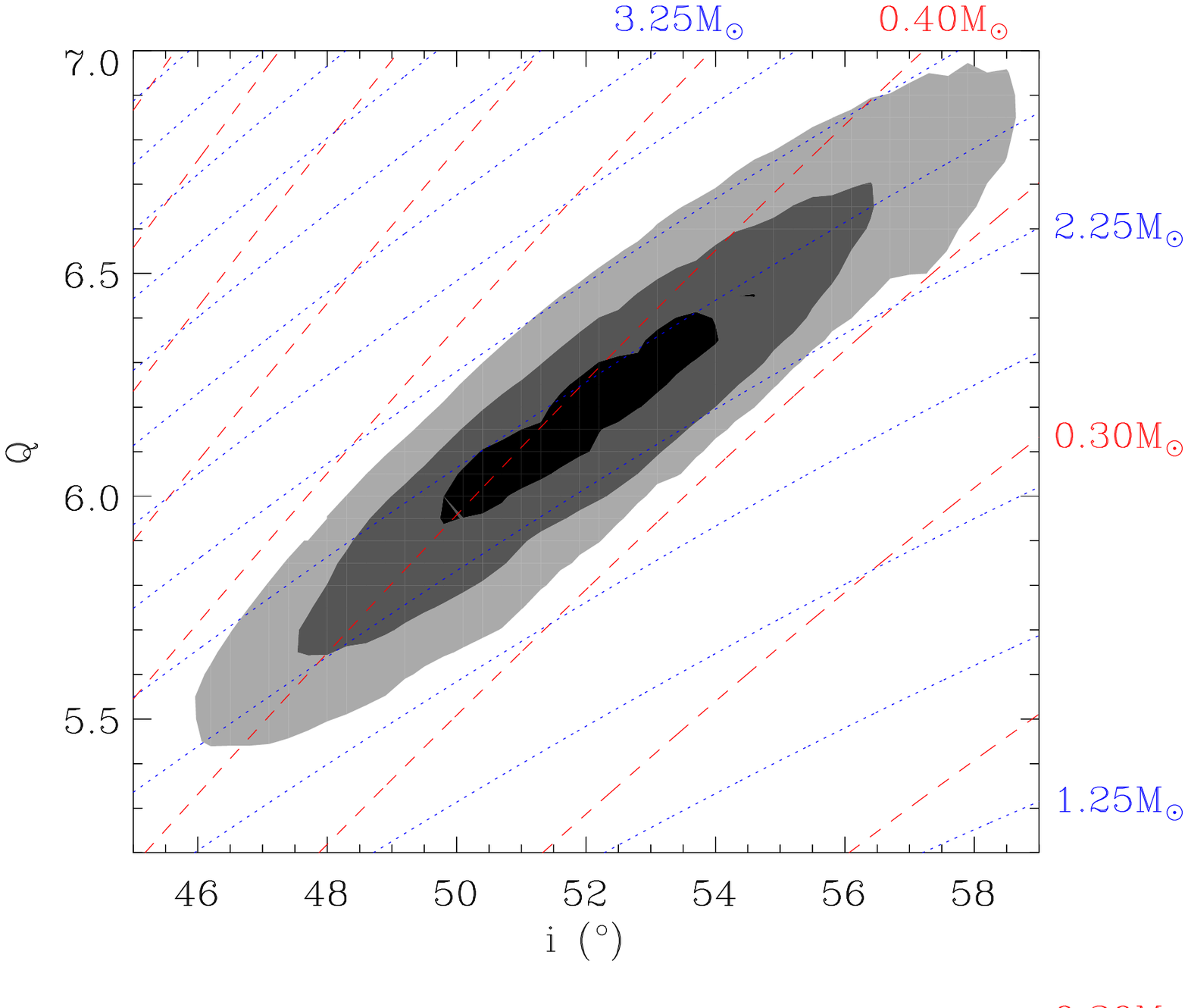}\tabularnewline
\includegraphics[scale=0.45]{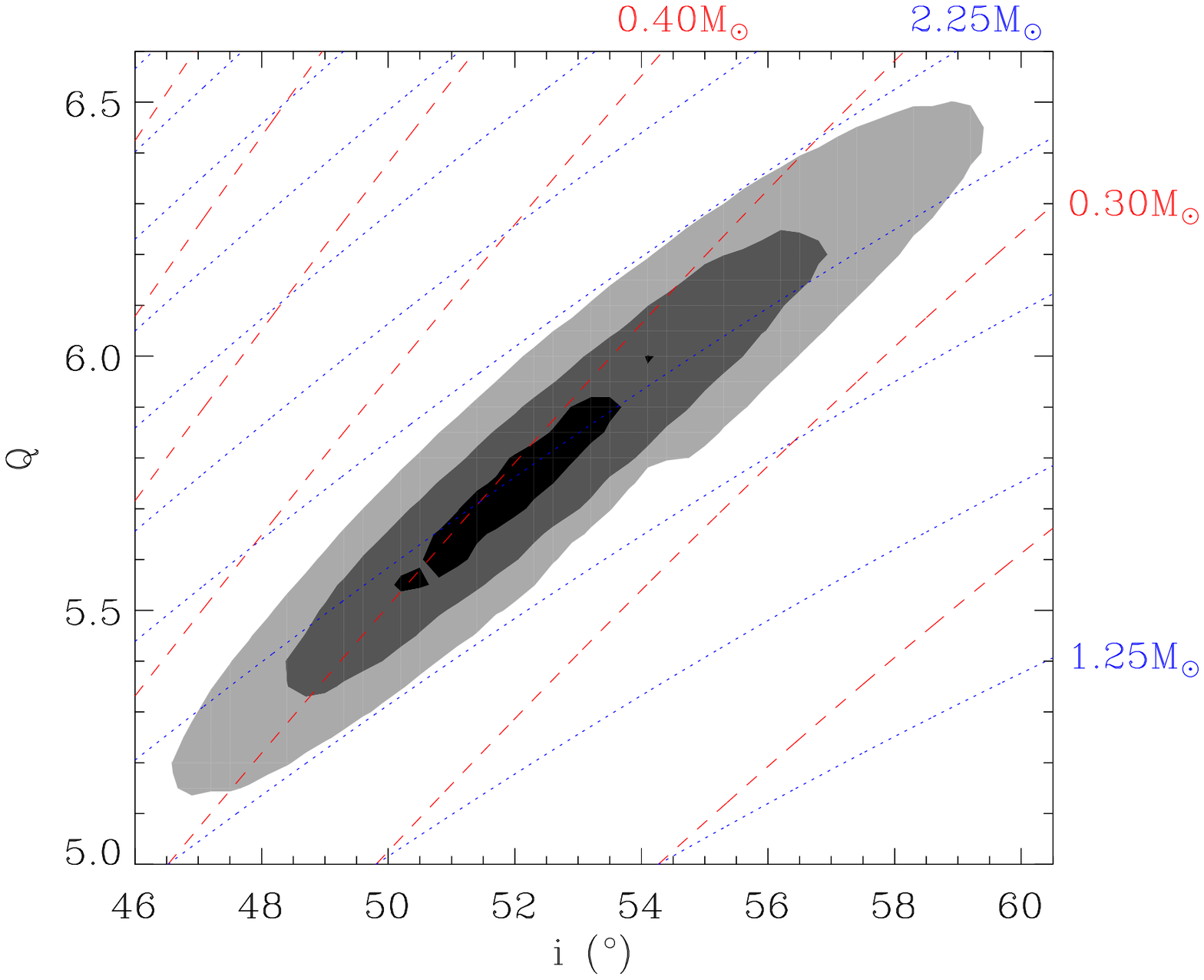}\tabularnewline 
\end{tabular}

\end{figure}

In contrast to J1810+1744, the counterpart to J2215+5135 is better-measured photometrically throughout its orbit, especially at minimum. This allowed us to fit the system parameters with greater certainty than for J1810+1744 and we find that the fits further are not complicated by implausible energetics or neutron star masses. As reported in Table~\ref{tab:2}, the uncertainties on the fit system parameters for both the NextGen and PHOENIX atmospheres (in both cases, reduced $\chi^2/\nu = 1.5$) are consistent, except for the $T_{\rm eff}$ which are only consistent to within their $\Delta \chi^2 = 9$ contours. This difference in $T_{\rm eff}$ causes a discrepancy in $T_{\rm hot}$ and corresponds to shifts in parameter space roughly orthogonal to the direction of lines constant mass in the $Q$ vs. $i$ planes plotted in Figure~\ref{fig:4}. Therefore, the derived parameters, $M_{\rm NS}$, $M_{\rm c}$ are also inconsistent to a similar degree while $K$ values for both models are consistent to within each other's $\Delta \chi^2 = 1.5$ contours.

As in the case of J1810+1744, the companion to J2215+5135 is found to be filling its Roche lobe to a high degree of certainty $\Delta f_{\rm Roche}/f_{\rm Roche} < 1 \times 10^{-3}$ due to the particular shape that its light curve exhibits with flattening at maximum and deepening at minimum being the most noticeable effects of the ellipsoidal variations. In this case, the precise shape is also enough to constrain the efficiency factor to a high degree of precision and it is consistent with the $\epsilon < 1$ limits. The phase shift measured for this system corresponds to the observed light curve leading the expected phase derived from the radio ephemeris by 144 seconds. We discuss this result in more detail in Section~\ref{sec:PhaseShifts}.

In effect, the only free parameters for this system are $i$, $Q$, and $T_{\rm eff}$, so we calculated the $\chi^2$ statistics around the best-fit values for these and then projected into the $Q$ vs. $i$ plane in Figure~\ref{fig:4} where we show the $\Delta \chi^2 = 1, 4,\ {\rm and}\ 9$ contours along with lines of constant inferred mass of the neutron star and the companion. The mass we find is comparable to that found in other black widow and redback pulsar systems \citep[e.g.]{Romani2012, vanKerkwijk2011}, with the unique feature that this determination is done without radial velocity data. This is possible because although the signal due to irradiation dominates the light curve observed, the particular shape of the light curve is different than what would be expected from a point-source illuminated spherical star. It is only by taking into account the ellipsoidal variations that the light curves can be fit and, indeed, the light curve is fit by a unique combination of $\epsilon$ and $f_{\rm Roche}$.

We note that the swing in color $B-V = 0.576$ to $B-V = 0.188$ corresponds to color temperatures of $T_{\rm color} = 6000 \pm 400\ {\rm K}$ to $T_{\rm color} = 7800 \pm 600\ {\rm K}$ when accounting for systematic uncertainty \citep{Flower1996} This temperature corresponds with the \citet{Breton2013} best-fits but is $\sim 1000\ {\rm K}$ hotter than the temperatures of the model fits reported in Table~\ref{tab:2}. ELC modeling of effective temperature is dependent only on the variation of flux in each individual band which are simultaneously fit without determining zero-points, and so are not dependent on the observed colors. If we instead assume that the measured color temperature is the minimum possible $T_{\rm eff}$ and rerun the model fitting procedure, we find an unreasonably large neutron star mass ($M_{\rm NS} > 3.6 M_\Sun$) at $\Delta \chi^2 = 100$. We take this as evidence that either the observed color is very aberrant or that there is an additional persistent and very blue source, for example a quiescent disk. The implications of this are discussed in more detail in Section~\ref{sec:Disk}.

The remaining free parameters quoted here are consistent with \citet{Breton2013}, including their inclination angle of $i = 66\degrees \pm16\degrees$ and filling factor of $f_{\rm Roche}=0.99\pm0.03$. They fit $\epsilon$ to a value approximately twice what we fit, but we note that if, as we might expect with redbacks, the bolometric albedo due to the reflection effect $a \sim 0.5$ is consistent with a convective envelope, our value matches their reported $L_{\rm x}/L_{\rm SD} \sim 0.15$.

\section{Discussion}
\label{sec:Discussion}

We now consider certain novel features of our results that have not been reported in the observations and modeling of similar systems. Specifically, we discuss 1) the dependence of constraints on ellipsoidal variations, 2) the phase shifts present in both objects, 3) the peculiarly high values of irradiance necessary to fit models of J1810+1744, and 4) the relatively blue colors of J2215+5135 as possible evidence for a quiescent disk in that system.

\subsection{Ellipsoidal Variations}

\begin{figure}[t!!]
\vskip 0.7truecm

\caption{\label{fig:5} For J1810+1744 (top) and J2215+5153 (bottom), solid curves are the expected signal due to the ellipsoidal variations calculated using the same parameters as in Table~\ref{tab:2}  except with $\epsilon = 0$. Plotted additionally are the residuals that are present if such a signal is removed from the model light curve in Figure~\ref{fig:2} with a zero point magnitude chosen to reasonably separate the light curves and the error bars propagated from the uncertainties in the flux difference. As in Figure~\ref{fig:2}, horizontal error bars are the exposure times scaled to the phase while the vertical error bars are the calculated magnitude uncertainties.}

\begin{tabular}{c}
\includegraphics[scale=0.55, trim=75 20 0 0]{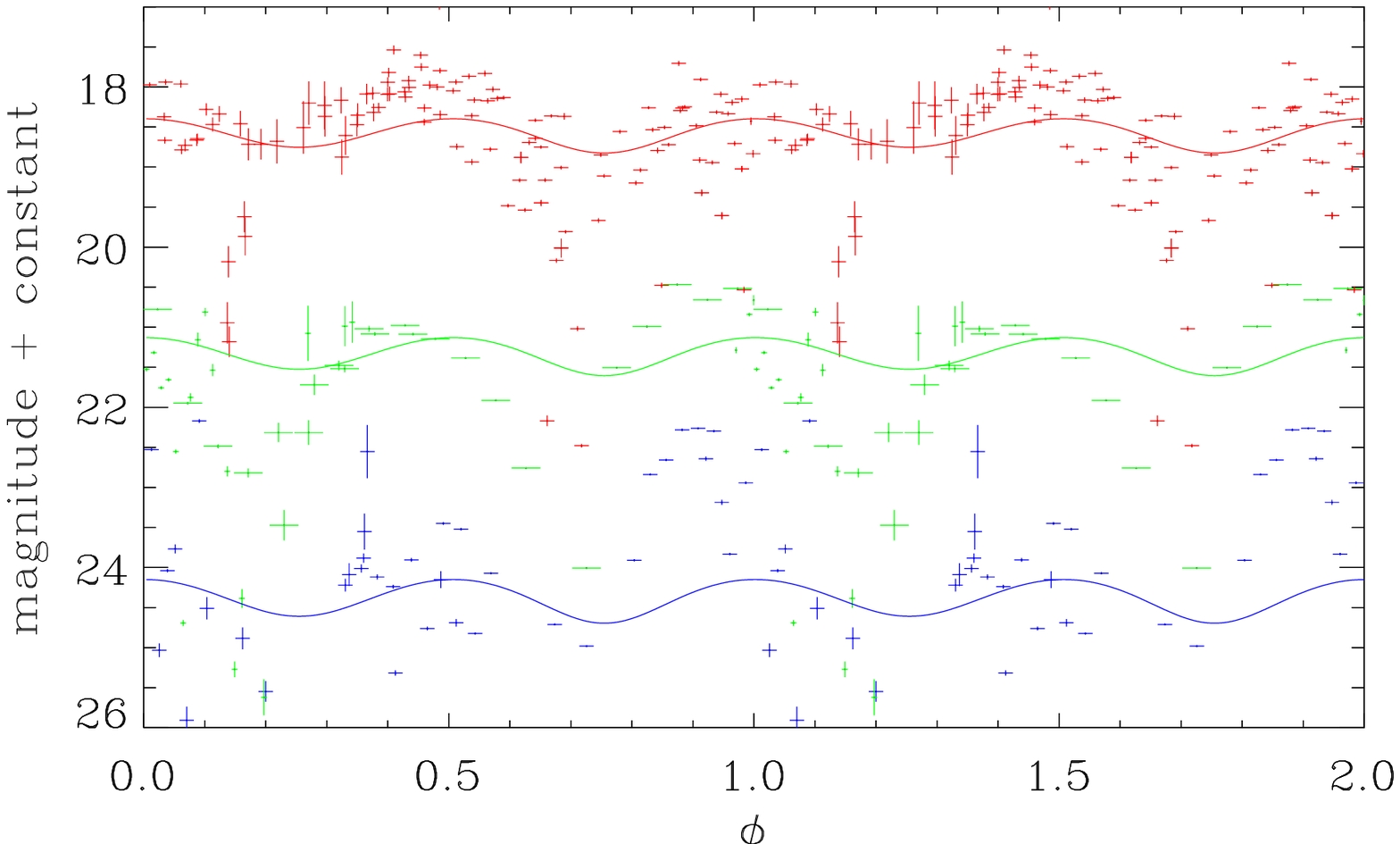}\tabularnewline
\includegraphics[scale=0.55, trim=75 20 0 0]{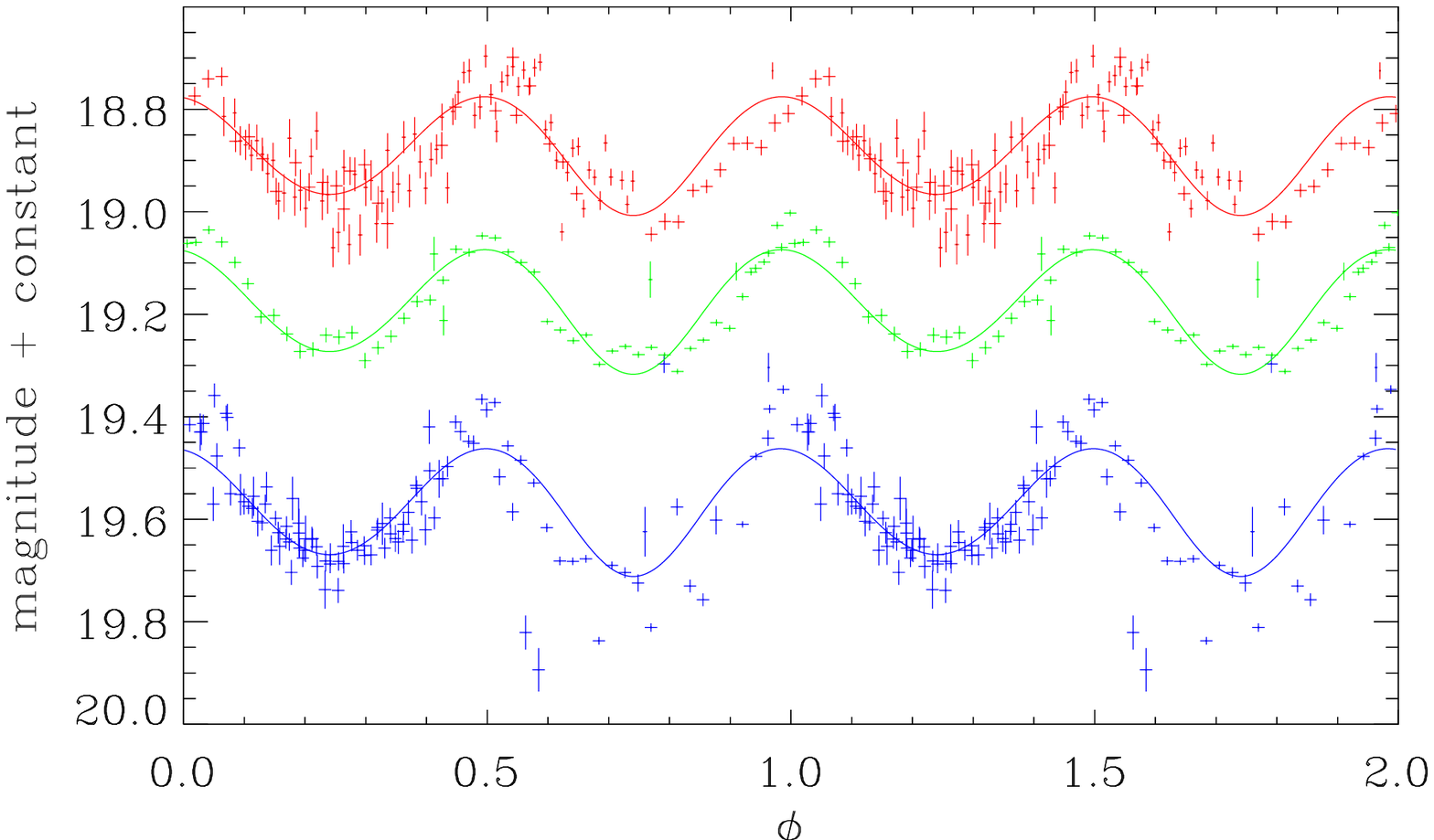}\tabularnewline 
\end{tabular}

\end{figure}

The model constraints we obtained for J2215+5135 were possible because the light curves were very well sampled and detailed features seen in them could be precisely fit. This can be seen because the ellipsoidal variations are detected in the light curves of these objects. We demonstrate this by comparing the signal due to irradiated heating with the signal due to the ellipsoidal variations seen in our observations by calculating the light curve of an non-irradiated system with the same $i$, $Q$, $T_{\rm eff}$, and $f_{\rm Roche}$ and then subtracting this from the model fit shown in Figure~\ref{fig:2} to obtain an ``irradiation only'' light curve. The flux associated with such an irradiation only light curve model is then be subtracted from the flux associated with our data to give the residuals which can be interpreted, roughly, as the amount of signal attributable to ellipsoidal variations alone. This, we show in Figure~\ref{fig:5}. For comparison's sake, we additionally plot in that figure the light curve of a similar non-irradiated system. To the extent the residuals do not match the ellipsoidal variations, the models may be failing to properly account for nonisotropic heating or a lack of orbit-to-orbit stability of the stellar flux at each phase.

Ellipsoidal variations in Roche lobe geometry are constrained to a level of $\sim 0.25$ mag in amplitude and the amplitudes decrease for smaller values of either $f_{\rm Roche}$ or $i$. In principle, the ellipsoidal variation minima at $\phi = 0.25$ and $\phi = 0.75$ can be fit by tuning $\epsilon$ and $T_{\rm eff}$ to larger values to accommodate smaller variations rather than adjusting other system parameters. In contrast, the ellipsoidal variation maxima ($\phi =0$ and $\phi=0.5$) coincides with the part of the light curve where the heating curve's gradient is largest and thus where the flux is changing most rapidly. The shape of the light curve at these phases therefore can strongly constrain the $f_{\rm Roche}$ and essentially breaks the degeneracy. This is the primary means by which the filling factor in these light curves is constrained and, as this value is set, so are the best-fit locations in $Q$, $i$, $\epsilon$ space for J1810+1744, and $Q$, $i$ space for J2215+5135 (where $\epsilon$ is strongly constrained by the less noisy signal at minimum). The fact that the ellipsoidal variations as well as the irradiation signal models are better fits to the data in J2215+5135 is confirmed in Figure~\ref{fig:5}. Even so, the variations in the residuals in J1810+1744 are phased with the ellipsoidal variations, and $f_{\rm Roche} = 1.0$ models are strongly preferred as the excess signal is actually larger than what is maximally possible.
 
\subsection{Phase Shifts}
\label{sec:PhaseShifts}

It has been suggested that the observed heating of black widow and redback systems could be attributed to either particle or radiation fields emanating from the stand-off shock between the neutron star and the companion \citep{Bogdanov2011}. The statistically significant phase shifts seen in our fits may imply that the sinusoidal heating signal is offset from phase alignment by 62 seconds late for J1810+1744 and 144 seconds early for J2215+5135. We interpret this to mean that there is an offset between the line connecting the pulsar and the longitude of the stand-off shock from whence the heating irradiance is emitted. Future modeling of the irradiation of short period binary millisecond pulsars should include both offsets and extended emission geometries given this and the issues outlined in Section~\ref{sec:J1810irradiation}.

While statistically significant in our model fits, the hypothesis that these shifts are due to offsets in source heating should be, in principle, testable if the ellipsoidal variations are observed because that signal should be in phase with the radio ephemeris measured $T_0$. Unfortunately, refitting the residuals in Figure~\ref{fig:5} did not result in a statistically significant difference in the phase shift, though the increased uncertainty in those fits does not rule out the hypothesis that the phase shift is due to heating offsets.

\subsection{Irradiation in J1810+1744}
\label{sec:J1810irradiation}

We find evidence that the received irradiation implied from the heating curve of J1810+1744 is in excess of the typical value given for the spin-down luminosity. According to our model fits presented in Figure~\ref{fig:3}, the best-fit luminosities lie between $4.0 < \epsilon < 10$ for models that allow for best-fit neutron star masses between $1.0 M_\Sun \leq M_{\rm NS} \leq 3.6 M_\Sun$. This is consistent with the best-fit models quoted by \citet{Breton2013} for this object.

While such efficiencies superficially seem to be in excess of what would be permitted given the energy budget afforded by the canonical spin-down luminosity calculated for J1810+1744, there are some straightforward arguments that could lead us to consider that a larger amount of incident radiation from the neutron star onto the companion can be accommodated. In particular, the assumption that the irradiation originating with the neutron star is isotropic is likely to be inaccurate. For example, MHD simulations of outflows \citep{Komissarov2004} and the characteristics of the termination or stand-off shock associated with pulsar winds \citep{Petri2011, Sironi2011} indicate that while the winds are likely axisymmetric, they are not likely to be spherically symmetric. If the energy released by the pulsar is preferentially emitted at a bright equatorial stand-off shock, the heating of the companion would be in excess of what would be naively calculated assuming an isotropic point source emanating from the neutron star.

\subsection{Evidence for a Quiescent Disk in J2215+5135}
\label{sec:Disk}

The tension between the color temperature observed for J2215+5135 and the model temperature we take as evidence for an additional source. Such a source would have to be comparable in luminosity to the companion star, but significantly hotter. Additionally, since the data indicate photometric stability even over many orbits, the discrepant source must be relatively stable in temperature and luminosity.

J2215+5135 is in a class of redback objects which compare to J1023+0038, the so-called ``missing link'' pulsar, that was observed to transition from an accretion-powered to a rotation-powered pulsar while also losing the emission features associated with its accretion disk \citep{Wang2009}. However, as accretion onto the neutron star ends, it is possible that a quiescent disk could remain between the companion and the light cylinder of the neutron star \citep{Eksl2005}.

Given $\eta = A_c / A_d$, the ratio of the visible surface area of the companion at phase $\phi = 0.75$ ($A_c$) to the visible surface area of any disk present in the system ($A_d$), a blackbody temperature for the disk is $T_d = \left[T_{\rm color}^4 + \eta \right(T_{\rm color}^4 - T_{\rm hot} \left)\right]^{1/4}.$  Assuming no reddening, $\eta = 1$ implies a disk temperature of $T_d = 9000\ {\rm K}$, while correcting for the full $E_{B-V} = 0.372$ reddening in that direction \citep{Schlegel1998} implies a temperature of $T_d = 19\ 000\ {\rm K}$.

The existence of such a disk in this system would imply that an expectation for similar behavior in J2215+5135 as that observed in J1023+0038 and IGR J18245-2452 with the pulsar alternating between accretion-powered and rotation-powered modes. If the object were to begin accreting, this would result in an increase in x-ray luminosity and the presence of strengthened emission lines.

\section{Conclusions}
\label{sec:Conclusions}

We present evidence that the photometric light curves observed from J2215+5135 is consistent with a neutron star of mass $2 M_\Sun$. This mass estimate based on photometry alone was possible because the light curve was sampled well enough to fit both the shape and amplitude of the heating and ellipsoidal variations. This result can be used as a prediction of what radial velocities we expect will be detected when this object is measured spectroscopically. Additionally, spectroscopic observations could be used to confirm the presence of a quiescent disk associated with this redback. If J2215+5135 is similar in type to J1023+0038, ongoing x-ray monitoring of the object will be useful in detecting if the object transitions from a rotation powered to an accretion powered system. This high-mass neutron star is readily comparable in size the measurements of \citet{Romani2012} and \citet{vanKerkwijk2011}. 

In contrast, J1810+1744, the second object for which detailed light curves were obtained, was not adequately constrained by our modeling. Whether this is due to a failure of the point-source illumination model or whether it may be due to intrinsic variability of the object is not clear. Spectroscopic observations of this object will be vital in constraining the mass ratio and therefore the neutron star mass in the system.

In addition, we report the first detection of the optical counterpart of J2214+3000 at $R = 22.64$ around the  the expected phase ($\phi = 0.75$) of maximal heating.

\medskip
\medskip

We thank Jerry Orsoz, Mark Reynolds, Saeqa Vrtilek, and Robin Barnard for helpful discussions and suggestions. 

\bibliographystyle{apj}
\bibliography{BWpulsar}

\end{document}